# EEG changes and motor deficits in Parkinson's disease patients: Correlation of motor scales and EEG power bands

Aleksandar Miladinović[a,b,*], Miloš Ajčević[b], Pierpaolo Busan[c,d], Joanna Jarmolowska[c], Manuela Deodato[d,e], Susanna Mezzarobba[e,f], Piero Paolo Battaglini[d], Agostino Accardo[b]

[a]*Institute for Maternal and Child Health – IRCCS Burlo Garofolo, Trieste, 34137, Italy*
[b]*Department of Engineering and Architecture at the University of Trieste, Trieste, 34127, Italy*
[c]*Fondazione Ospedale San Camillo IRCCS, via Alberoni 70, 30126 Venice, Italy*
[d]*Department of Life Sciences at the University of Trieste, Trieste, 34127, Italy*
[e]*Department of Medicine, Surgery and Health Sciences at the University of Trieste, Trieste, 34149 Italy*
[f]*Department of Neurosciences, Rehabilitation, Ophthalmology, Genetic and Maternal and Infantile Sciences, University of Genova, Genova, 16132, Italy.*

**Abstract**

Over the years motor deficit in Parkinson's Disease (PD) patients was largely studied, however, no consistent pattern of relations between quantitative electroencephalography (qEEG) and motor scales emerged. There is a general lack of information on the relation between EEG changes and scales related to specific motor deficits. Therefore, the study aimed to investigate the relation between brain oscillatory activity alterations (EEG power bands) and most used PD-related motor deficit scales. A positive correlation was found between the *freezing of the gait questionnaire* (FOGQ) and delta spectral power band *(rho=0.67; p=0.008),* while a negative correlation with the same scale was observed in the alpha spectral power band *(rho=-0.59, p=0.027).* Additionally, motor scores measure by motor part of *Unified Parkinson's Disease Rating Scale* (UPDRS) correlated directly with theta *(rho=0.55, p=0.040)* and inversely with beta EEG power band *(rho=-0.77, p=0.001).* No significant correlation was found between spectral powers and *Hoehn and Yahr* (H&Y), *BERG* (Berg K. et. al. 1995), *Modified Parkinson Activity Scale* (MPAS), *Six-Minute Walk Test* (6MWT) and *Timed Up and Go Test* (TUG). In conclusion, our study supports the earlier findings suggesting a link between EEG slowing and motor decline, providing more insight into the relation between EEG alteration and deficits in different motor domains. These findings indicate that EEG assessment may be a useful biomarker for objective monitoring of progression and neurophysiological effect of rehabilitation approaches in PD's.

* Corresponding author. Tel.: +39 040 558 7130
E-mail address: aleksandar.miladinovic@burlo.trieste.it

| **Nomenclature** | |
|---|---|
| PD | Parkinson's Disease |
| EEG | Electroencephalography |
| qEEG | Quantitative EEG |
| H&Y | Hoehn and Yahr |
| FOG | Freezing of Gait |
| FOGQ | FOG Questionnaire |
| UPDRS-III | Unified Parkinson's Disease Rating Scale Part III (motor examination) |
| 6MWT | Six-Minute Walk Test |
| TUG | Timed Up and Go Test |
| MPAS | Modified Parkinson Activity Scale |

## 1. Introduction

Parkinson's Disease (PD) is a neurodegenerative disorder characterized by motor and cognitive impairments coupled with an alteration of the electrophysiological oscillatory activity. Around 70% of PD patients experience tremors during the disease [1]. These primary motor symptoms are usually followed by difficulties in initiating movements, gait disturbance and speech problems drastically affecting patients' quality of life [2].

Various scales have been developed for the quantification of motor abilities of PD patients. The UPDRS has achieved the greatest acceptance as the main clinical assessment and progression monitoring tool [3]. However, there are other supplementary scales that focus on specific aspects of motor deficits and disease development, such as the Hoehn and Yahr (H&Y) scale [4], which assesses the patterns of progressive motor decline. Likewise, the walking capabilities of the patients can be quantified by 6 Minutes Walking Test (6MWT) [5] and by Timed Up & Go (TUG) test that measures functional mobility and gait speed [6]. In addition, a specific scale that assesses Freezing of Gait (FOG), one of the most disabling symptoms in PD patients, is usually assessed by a specifically designed questionnaire (FOGQ) [7,8].

There is a growing research interest towards the neurophysiological biomarkers for monitoring the progression of neurological disability usually measured by clinical scales [9–14]. Currently, there are no reliable objective biomarkers for disease progression in PD [9]. Quantitative biomarkers may identify systems at risk before the overt expression of the disorder. Biomarkers ideally should be techniques that are non-invasive, widely available and economic. Electroencephalography (EEG), as a technique that combines the aforementioned aspects, provides insight into cortical function by measuring cerebral electrical activity and by quantitative analyses of the brain can provide more information about the cortical dysfunction [12,15,16].

Changes in oscillatory brain activity in the relative powers of the delta, theta, alpha and beta bands have been observed in PD patients during resting-state EEG recordings. In particular, alpha and beta bands show a decrease in the relative powers, even in the initial phase of the disease, while delta and theta bands display an increase in their relative powers [17–19]. These findings may represent a breakthrough in the research, as those brain wave changes may represent the hallmark of EEG abnormalities in PD that can be important to better understand the disease, monitor its progress and evaluate treatment efficacy. Furthermore, the design of Brain-Computer Interface motor rehabilitation [20], feasible in neurologic disorders [21,22], and in control of assistive robots [23] could be improved by understanding EEG changes in PDs. Indeed, EEG changes over time can introduce nonstationarities and diminish BCI accuracy [24]. The intersession transfer learning techniques improve the performance in PDs [25], but its further advance requires a better understanding of the PD pathology related EEG.

The correlations between EEG slowing and global cognitive impairment in PDs were widely studied [11,26–30], and the papers are reporting consistent findings. On the other hand, a few studies investigated the relation between EEG alterations and motor deficits, focusing mainly on various revisions of the Unified Parkinson's Disease Rating Scale (UPDRS) and Hoehn & Yahr (H&Y) scale [9,10,29,31], and therefore, there is a lack of information on the relation of EEG changes and with scales related to specific motor domains.

Hence, this study aimed to investigate the relation between brain oscillatory activity alterations and most used PD-related motor deficit scales.

## 2. Materials and Methods

*2.1. Participants and Study protocol*

Seven PD patients (4M/3F; age 72±4.5 years) were enrolled in the study. The inclusion criteria were: diagnosis of PD; diagnosis of gait's FOG disturbance; stage of Hoehn and Yahr Scale < 3; cognitive Mini-Mental State Examination (MMSE) score > 24; stable pharmacological treatment for at least two months. All included patients gave their signed consent for participation in the study. The experimental protocol was pre-approved by the Local Ethical Committee and was conducted according to the principles of the Declaration of Helsinki.
All patients underwent motor deficit assessment made up of a battery of motor tests followed by the acquisition of the EEG. In particular, UPDRS-III, H&Y, 6MWT, Berg, TUG, FOGQ motor tests and questionnaires were performed and corresponding scores were assigned. The relations between relative power for each spectral band and the aforementioned motors scales were investigated. All the evaluations were conducted in the pharmacological "on" state of PD patients.

*2.2. Motor scales*

**Unified Parkinson's Disease Rating (UPDRS)**

The Unified Parkinson Disease Rating Scale (UPDRS) is a tool to judge the course of the disease in the patients and consists of 5 segments: 1) Mentation, Behavior, and Mood, 2) Activities of Daily Living (ADL), 3) Motor section, 4) Modified Hoehn and Yahr Scale and 5) Schwab and England ADL scale. In this study, we considered the third, motor-related, section of the overall UPDRS. UPDRS-III consists of 14 items aimed at evaluating the speech abilities, the resting tremor, the postural stability, the rigidity, the hand movements, bradykinesia and hypokinesia of the PD patients, assigning a score from 0 to 4 points for each item depending on the degree of impairment.

**Hoehn and Yahr (H&Y)**

The Hoehn and Yahr (H&Y) scale is a clinical rating scale that captures the patterns of progressive motor decline. However, because of its simplicity and lack of details, it is limited to some aspects of PD disease, such as the presence or absence of postural reflex impairment, leaving other aspects of the motor deficit untreated. The H&Y scale consists of 1 to 5 stages (Table 1) where the last two indicate a severe disability and less independence of the patient [32].

Table 1. Hoehn and Yahr scale Stage Description

| Stage | Symptoms |
|---|---|
| 1 | Unilateral involvement only, usually with minimal or no functional disability |
| 2 | Bilateral or midline involvement without impairment of balance |
| 3 | Bilateral disease: mild to moderate disability with impaired postural reflexes |
| 4 | Severely disabling disease; still able to walk or stand unassisted |
| 5 | Confinement to bed or wheelchair unless aided |

**Modified Parkinson Activity Scale (MPAS)**

MPAS consists of 14 items organized into 3 domains as chair transfer, gait akinesia, and bed mobility. Also, this scale uses a 5-points scoring system from 0 to 4 with a total score ranging from 0 (best) to 56 (worst) performance [33].

**Freezing of Gait Symptom Questionnaire (FOGQ)**

FOG severity in maximum 10 minutes: it consists of 6 items and a 5-points scale ranging from 0 (absence of symptoms) to 4 (most severe) for a total score of 24, where higher scores correspond to most severe FOG [7,8].

**Six-Minute Walk Test (6MWT)**

Among different symptoms, reduced velocity is a well-defined characteristic of PD patients. It is possible to provide a measure of the walking capacities of the patients by using the 6 Minutes Walking Test (6MWT) which typically combines numerous turns and straight-line walking to quantify moving abilities by measuring the walked distance in 6 minutes.

**Timed Up & Go (TUG)**

Finally, the Timed Up & Go (TUG) test correlates with functional mobility, gait speed and falls as it measures the time taken by a subject to (1) rise from an ordinary chair (without arms), (2) walk 3 meters, (3) turn for 180°, (4) walk back to the starting position and (5) sit down. The process should be performed as fast as possible. The shorter the time to perform the activity, the better the mobility of the patient is. Concerning PD patients, TUG cut-off scores range from 8 to 11.5 seconds, and longer TUG times are associated with decreased mobility and with increased risk of falls [34].

*2.3. EEG acquisition and processing*

EEG signals were acquired by using SAM 32FO amplifier (Micromed, Italy) with a sampling rate of 256Hz and a prewired head cap with 23 Ag/AgCl electrodes (Electro-Cap International, Eaton, OH, USA) placed at standard 10-20 (Fp1, Fpz, Fp2, F7, F3, Fz, F4, F8, T3, C3, Cz, C4, T4, T5, P3, Pz, P4, T6, O1, Oz, O2). The reference electrode was placed in POz, while the ground electrode was placed in AFz. Electrode impedances were kept below 5 kΩ.
EEG offline analysis of the resting state data was carried out using MATLAB(R) (The MathWorks Inc., Natick, MA). All channels were filtered 0.5 - 45 Hz with the 2nd order Butterworth bandpass filter. Artefacts were manually rejected after the visual inspection of tracings. The Power Spectral Density (PSD) was extracted on 80 seconds long data by using the approach based on Welch's periodogram [31] (averaged on 15 tracts of 10 s each, windowed with a Hann window, with 50% overlap). The band powers: delta (1-4Hz), theta (4-8 Hz), alpha (8-13 Hz), beta (13-30 Hz) were calculated and normalized with the total power in the range 1 to 30 Hz. The relative power for each band was averaged over all 21 channels.

*2.4. Statistical analysis*

Continuous variables with a normal distribution are presented as means and standard deviations, those with a skewed distribution as median and ranges. The correlation between the relative power of each spectral band and assessed motor deficit scores were investigated using Spearman's nonparametric test. The value of $p < 0.05$ was considered significant.

**3. Results**

The median (range) values obtained from motor assessment data of seven enrolled PD patients are summarized in Table 2. The median relative delta, theta, alpha and beta power were 0.12 (range 0.04–0.16), 0.21 (range 0.13–0.45), 0.25 (range 0.12–0.44) and 0.23 (range 0.13– 0.35), respectively. Results of correlation between relative powers of each spectral band with Motor Scales are reported in Table 3. The scale which expresses the deficit related to freezing of gait symptom (FOGQ) correlated significantly with delta and alpha band. In particular, a positive correlation was found with delta (rho=0.67; p=0.008), while a negative correlation was observed with alpha (rho=-0.59,

p=0.027). Moreover, the motor-related third section of commonly used UPRS (i.e UPDRS-III) correlated directly with theta (rho=0.55, p=0.040) and inversely with beta (rho=-0.77, p=0.001). No significant correlation was found between spectral powers and H&Y, BERG, MPAS, 6MWT and TUG scales.

**Table 2**. The median (range) values obtained from motor assessment data of enrolled PD patients.

| Motor scale | Median (range) |
|---|---|
| **FOGQ** | 6 (0 - 11) |
| **H&Y** | 1.5 (1.0 - 2.5) |
| **UPDRS-III** | 8.5 (4.0 - 24.0) |
| **BERG** | 52.5 (45.0 - 58.0) |
| **MPAS** | 60 (53 - 64) |
| **6MWT** | 360 (262 - 461) |
| **TUG** | 909.5 (753.0 - 1366.0) |

**Table 3**. Correlation between relative powers of each spectral band and motor scales (significant results are marked with boldface and asterisk).

| | Spearman's rho (p-value) | | | |
|---|---|---|---|---|
| | *delta* | *theta* | *alpha* | *beta* |
| **FOGQ** | **0.67 (p=0.008) *** | 0.33 (p=0.224) | **-0.59 (p=0.027) *** | -0.39 (p=0.169) |
| **H&Y** | -0.02 (p=0.945) | 0.17 (p=0.550) | 0.19 (p=0.524) | -0.44 (p=0.113) |
| **UPDRS** | -0.03 (p=0.909) | **0.55 (p=0.040) *** | 0.09 (p=0.766) | **-0.77 (p=0.001) *** |
| **BERG** | 0.24 (p=0.415) | -0.38 (p=0.182) | 0.08 (p=0.798) | 0.23 (p=0.428) |
| **MPAS** | 0.26 (p=0.370) | 0.05 (p=0.872) | -0.06 (p=0.848) | -0.16 (p=0.573) |
| **6MWT** | 0.45 (p=0.106) | -0.20 (p=0.502) | -0.36 (p=0.205) | 0.37 (p=0.188) |
| **TUG** | -0.42 (p=0.132) | 0.02 (p=0.952) | 0.46 (p=0.097) | -0.25 (p=0.382) |

## 4. Discussion

The motor deficit in PD's was largely studied, however, over the years, no consistent pattern of relations emerged between qEEG variables (spectral power features) and scales of the motor deficits [9]. Therefore, the focus of the work was to investigate the association of EEG spectral bands with a decline of different motor domains in patients with PD.

The main finding of this study is a significant positive correlation delta and negative correlation with alpha-band freezing of gait related FOGQ scale. This is the first study that assessed the correlation of these EEG alterations with an increased number of FOG episodes. The observed effect can be explained by the presence of general EEG slowing (an increase of power in lower frequency bands and increase of power in higher frequency bands) in PDs as the disease progresses [26–30,35], affecting one's ability to initiate and coordinate motor movements. We have also found a similar trend between high and low-frequency bands and the motor part of the UPDRS scale. In this case, the positive correlation was observed in the theta band and negative in the beta band which is in line with the recent study [29] and, furthermore, with a study on early-onset PD that also reports beta band power coherence decrease with the deterioration measured by MDS-UPDRS III [36].

Studies have reported alternation of the resting-state EEG as power increase of lower frequency bands (delta and theta) and decrease in higher frequency bands (beta, gamma), also called PD patients EEG slowing [17–19]. Correlation of various scales that assess cognitive and motor dysfunction is required to better understand the cause of the observed EEG changes. In contrast to the cognitive domain where many different scales were compared with EEG, the motor domain remained limited to a comparison of UDPRS and H&Y. The published results are, furthermore, inconsistent, and requires broader investigation involving different motor aspects [9]. The inconsistency in the results reported in the former studies might be due to pharmacological treatment. Some studies [36,37] report correlations of alpha and beta powers and UPDRS induced by Levodopa administration. Furthermore, some types of non-pharmacological treatments, such as BCI Motor-Imagery, Action Observation or their combination with standard therapeutic practice can also alter the resting-state EEG of subjects [38–40]. Despite the small sample size, our study supports the earlier findings suggesting a link between EEG slowing and motor decline, providing more insight into the relation between EEG alteration and single motor domains. Future investigations with larger sample sizes and control medication administration are needed to confirm clinical aspects of mutual interchangeability of EEG power band slowing and motor impairment in PD patients.

## 5. Conclusions

In this study, we identified the association between qEEG changes and different motor deficit domains measured by specific motor scales. The significant correlation between EEG slowing and symptom-specific motor decline indicates that EEG assessment may be a useful biomarker for objective monitoring of progression and neurophysiological effect of rehabilitation approaches in PD's.

## Acknowledgements

Work partially supported by the master programme in Clinical Engineering of the University of Trieste.

## Conflict of Interest Statement

The authors declare that the research was conducted in the absence of any commercial or financial relationships that could be construed as a potential conflict of interest